\documentclass{emulateapj}
\usepackage{apjfonts}

\shorttitle{PKS~2155-304}
\shortauthors{Fang et al.}

\makeatletter

\begin{document}

\title{Confirming the Detection of an Intergalactic X-ray Absorber
  Toward PKS~2155-304} 

\author{Taotao~Fang\altaffilmark{1},
        Claude~R.~Canizares\altaffilmark{2}, Yangsen~Yao\altaffilmark{2}}

\altaffiltext{1}{Department of Physics and Astronomy, University of California,
        Irvine, CA 92697, fangt@uci.edu; {\sl Chandra} Fellow}
\altaffiltext{2}{Department of Physics and Center for Space Research,
        MIT,  77 Mass. Ave., Cambridge, MA 02139} 

\begin{abstract}

We present new observations on PKS 2155-304 with the {\sl Chandra} Low
Energy Transmission Grating Spectrometer (LETG), using the Advanced
CCD Imaging Spectrometer (ACIS). We confirm the detection of an
absorption line plausibly identified as \ion{O}{8} Ly$\alpha$ from the warm-hot intergalactic medium associated with a small group of
galaxies along the line of sight, as originally reported by Fang et
al.~2002 (here after FANG02). Combining the previous observations in
FANG02 and five new, long observations on the same target, we increase
the total exposure time by a factor of three, and the total counts per
resolution element by a factor of five. The measured line equivalent
width is smaller than that observed in FANG02, but still consistent at
90\% confidence. We also analyze the {\sl XMM}-Newton observations on
the same target, as well as observations using the {\sl Chandra} LETG
and the High Resolution Camera (HRC) combination. These observations
have been used to challenge our reported detection. While no line is
seen in either the {\sl XMM}-Newton and the {\sl Chandra} LETG+HRC
data, we find that our result is consistent with the upper limits from
both data sets. We attribute the non-detection to (1) higher quality
of the {\sl Chandra} LETG+ACIS spectrum, and (2) the rather extended
wings of the line spread functions of both the {\sl XMM}  RGS and the
{\sl Chandra} LETG+HRC. We discuss the implication of our observation
on the temperature and density of the absorber. We also confirm the
detection of $z\sim 0$ \ion{O}{7} absorption and, comparing with
previous {\sl Chandra} analysis, we obtain much tighter constraints on
the line properties.

\end{abstract}


\section{Introduction}

Cosmological hydrodynamic simulations predict that a large amount of
baryons in the local universe reside in the intergalactic medium (IGM)
(see, e.g., Cen \& Ostriker~1999; Dave et al.~2001; Kravtsov et al.~2002; Cen \&
Ostriker~2006). While the warm
and photo-ionized Ly$\alpha$ clouds contain a significant
fraction of the total baryons (see, e.g., Morris et al.~1991; Bahcall et al.~1991;Stocke et al.~1995;Shull et al.~1996;Penton et
al.~2000; Williger et al.~2006;Lehner et al.~2007), the
remainder is shock-heated to temperatures between $10^5$
-- $10^7$ K with moderate overdensities. Highly ionized
metals in this Warm-Hot Intergalactic Medium, or ``WHIM'', can produce
detectable absorption signatures in the UV/X-ray spectra of background sources,
just as the cooler IGM imprints the Ly$\alpha$ forest (see, e.g.,
Perna \& Loeb~1998;
Hellsten~1998; Fang, Bryan, \& Canizares~2002; Chen et al.~2002; Viel
et al.~2003; Fujimoto et al.~2004; Furlanetto et al.2005; Cen \& Fang ~2006;Kawahara et al.~2006). Detectability of the WHIM gas through emission is also under investigation (see,
e.g., Croft et al.~2001; Phillips et al.~2001; Kuntz \& Snowden et al.~2001;
Zappacosta et al.~2002, 2005,2007; Yoshikawa et al.~2003, 2004;
Kaastra~2004; Mittaz et al.~2004; Slotan et al.~2005;Fang et al.~2005; Ursino \&
Galeazzi~2006;Cen \& Fang~2006; Takei et al.~2007a,2007b; Mannucci et al.~2007). 

Recent detections of the UV absorption lines with the STIS onboard the
{\sl Hubble} Space Telescope and the Far Ultraviolet Spectrometer
({\sl FUSE}) have firmly revealed the existence of the WHIM gas at
temperatures between $10^5$ -- $10^{5.5}$ K (see, e.g., Savage et
al.~1998; Tripp et
al.~2000; Oegerle et al.~2000; Sembach et al.~2004; Tumlinson et
al.~2005; Danforth \& Shull~2005; Stocke et al.~2006). The \ion{O}{6}
doublet at 1032 and 1039 \AA\, can be used to probe the 
temperatures, densities and redshifts of that component of the
intervening WHIM gas. Recently, a population of intervening broad HI
Ly$\alpha$ absorbers (BLAs) have also been detected in the far-UV band,
stirring interests in probing WHIM with BLAs (see. e.g., Richter et
al.~2005; 2006). The contribution of both BLAs and \ion{O}{6} absorbers to the baryonic density are $\Omega_b(\rm BLA) \geq 0.0027$ (Richter et al.~2005) and $\Omega_b(\rm O VI) \sim 0.0022$ (see, e.g., Danforth \& Shull~2005), respectively, assuming a Hubble constant of $\rm 70\ km\ s^{-1}Mpc^{-1}$. However, as simulations suggest, about only one third of the WHIM gas can be detected in the UV band --- with higher temperatures the remaining two thirds can only be revealed in the
X-ray (see, e.g., Fang, Bryan, \& Canizares~2002; Chen et al.~2002;
Viel et al.~2003; Cen \& Fang~2006; Cen \& Ostriker~2006).

Several lines of evidence indicate that the narrow X-ray absorption
lines from the WHIM gas may be detected with high resolution
spectrometers onboard {\sl Chandra} and {\sl XMM}-Newton. A number of
$z \approx 0$ absorption lines were detected in the spectra of
background AGNs with {\sl Chandra} and
{\sl XMM}-Newton (Nicastro et al.~2002; Fang, Sembach, \& Canizares~2003;
Rasmussen et al.~2003; Kaspi et al.~2002;McKernan et al.~2004,2005;
Cagnoni et al.~2004; Williams et al.~2005, 2006a, 2006b). However, current instrumental resolution cannot
distinguish between Galactic and the Local Group origin of the
absorbing gas. While a number of studies proposed these absorption
lines are produced by the intragroup medium in the Local Group (see,
e.g.,  Nicastro et al.~2002; Williams et al.~2005, 2006a,2006b), other
evidence suggests a Galactic-origin of this hot gas (Wang et al.~2005;
Yao \& Wang~2005; Fang et al.~2006; Yao \& Wang
~2007; Bregman \& Lloyd-Davies~2007). 

Fang et al.~(2002, hereafter FANG02) reported the first detection
of an intervening X-ray absorber at $z = 0.0554$ along the sightline
toward PKS~2155-304, using the {\sl Chandra} LETGS. At the same
redshift a small group of galaxies had been detected by Shull et
al.~(1998), who subsequently reported the detection of \ion{O}{6}
(Shull et al.~2003). Subsequent observations with {\sl XMM}-Newton did not
detect the \ion{O}{8} absorption, setting a 3$\sigma$ upper limit
equivalent width (EW) of $\sim 14$ m\AA\, compared to our reported EW
of $14.0^{+7.3}_{-5.6}$ m\AA\ (Cagnoni et al.~2004). Several
\ion{O}{7} and \ion{O}{8} absorption lines were detected along the
sightline toward H~1821+621 (Mathur et al.~2003),  although the
statistics is low and the detections are at 2 -- 3$\sigma$
level. Mckernan et al.~(2003) claimed detection of an intervening
absorption system at $z \approx 0.0147$ toward 3C~120 at $\gtrsim
3\sigma$ level, although they cannot rule out the possibility that
this system is intrinsic to the jet of 3C~120. Recently, during an
extremely bright state of Mkn~421, Nicastro et al.~(2005) detected two
intervening absorption systems ($z>0$) with high significance. However, observations with the {\sl XMM}-Newton cannot
confirm the detections, and the consistency between the {\sl Chandra}
and {\sl XMM}-Newton was investigated (Ravasio et al.~2005; Williams et al.~2006c; Kasstra
et al.~2007; Rasmussen et al.~2007).   
 
In this paper, we confirm detection of the intervening absorber toward PKS~2155-304 using new observations to reexamine the results from FANG02. PKS~2155-304 has been used repeatedly as a
{\sl Chandra} and {\sl XMM}-Newton calibration target because it is
one of the brightest extragalactic soft X-ray sources, and also
because of its relatively simple spectrum shape. Since we published
FANG02, five more {\sl Chandra} observations with the same
instrumental configuration (LETG+ACIS) were conducted, which more than
triples the total exposure time that was reported in FANG02. The
increase in the number of detected photons (by a factor of $\sim 5$)
improves the statistics of the spectrum significantly. We also analyze
the {\sl XMM}-Newton observations on the same target, as well as
observations using the {\sl Chandra} LETG+HRC combination. These
observations have been used to challenge our reported detection. 

\section{Data Analysis}

At $z = 0.112$, PKS~2155-304
($\alpha=21^h58^m52.1^s,\,\delta=-30\arcdeg13\arcmin32.1\arcsec$) is
one of the brightest extragalactic X-ray sources. This source has been
chosen to calibrate various instruments onboard {\sl Chandra} and so
is observed repeatedly. For our
purpose, we choose observations that have been conducted with the Low
Energy Transmission Gratings (LETG) and ACIS-S as the focal plane
detector \footnote{For LETG and ACIS-S, see http://asc.harvard.edu}, to be consistent with FANG02. In section \S4, we will discuss results from observations using LETG with HRC-S as the focal plane detector. 

In FANG02, we report the results from three observations that were
conducted during May and December of 2000 and November of 2001, with a
total exposure time of 86.7 {\it ksec}. Subsequently, seven more
observations were conducted between June 2002 and September 2005. We
select those observations with at least 30 $ksec$ exposure time, which
results in a total of five more observations. Including these five
more observations increases the total exposure time to $\sim 277$ {\it
  ksec}, and the number of counts in the spectral regions of interests
is also increased by a factor of $\sim 5$. Table~1 lists the
observation log.  Seven of the eight observations have the nominal
offset pointing for LETG+ACIS-S: the aimpoint is moved by $\Delta
y=+1.5\arcmin$ along the observatory $y$ direction so that the most
interested wavelength range (0 -- 26 \AA) can be covered entirely by
backside chip S3, which has higher quantum efficiency and large
effective area, and the Scientific Instrument Module (SIM) is moved by
($\rm SIM-Z=-8 mm$), to mitigate the CTI (Charge Transfer Inefficiency)-induced energy resolution degradation. This also avoids putting the zeroth-order at one of the node boundaries. The only exception is Obs.\#3669, which had a $y$-offset of $3.3\arcmin$. This put the aimpoint on chip S2. 

\begin{table}[t]
\small
\caption{Observation Log}
\begin{center}
\begin{tabular}{cccc}
\hline \hline 
Observation ID & Observation Date & Duration (ksec) & Reference \\

\hline

1703 & 31~May~2000 & 26.7 & 1,2  \\
2335 & 06~Dec~2000 & 30.0 & 1,2  \\
3168 & 31~Nov~2001 & 30.0 & 1,2  \\
3669 & 11~Jun~2002 & 50.0 & 2  \\
3707 & 30~Nov~2002 & 30.0 & 2  \\
4416 & 16~Dec~2003 & 50.0 & 2  \\
6090 & 25~May~2005 & 30.0 & 2  \\
6091 & 19~Sep~2005 & 30.0 & 2  \\
\hline
\end{tabular}
\tablerefs{1. Fang et al.~(2002); 2. this paper.}
\end{center}
\end{table}

Data analysis is performed as described in FANG02. We briefly
summarize here and refer the reader to that paper for more details. The
{\sl Chandra} LETGS produces a zeroth order image at the aim-point on
the focal plane detector, the ACIS-S array, with higher order spectra
dispersed to either side. The LETGS provides nearly constant spectral
resolution ($\Delta\lambda = 0.05 \AA$) through the entire bandpass
(0.3-5 keV). The moderate energy resolution of the CCD detector ACIS-S
is used to separate the overlapping orders of the dispersed
spectrum. We add the plus and minus sides to obtain the first order
spectrum. All the five data sets are analyzed with the standard {\sl
Chandra} Interactive Analysis of Observations (CIAO) \footnote{See
http://asc.harvard.edu/ciao} and customized Interactive Data Language
(IDL) routines.

\section{Continuum Spectral Analysis}

The continua are typically fitted with a power law modified by
neutral hydrogen absorption in the range between
6 and 42 \AA.\ For observations \#1703, 2335, and 3168 we
refer to FANG02 for detailed model parameters. For \#3669, the data is
best fitted with a broken power law. The photon
indices are $2.52\pm0.02$ and $1.96\pm0.02$, with a break energy of
$\sim 2$ keV, and a flux of
$1.66\times 10^{-10}\rm\ ergs\,cm^{-2}s^{-1}$ between 0.5 and 2.4 keV. The
remaining four observations can be best fitted by a single power law
with photon indices of $(2.69\pm0.02, 2.69\pm0.02, 2.58\pm0.01,
2.66\pm0.01)$, for \#3707, 4416, 6090, and 6091, respectively. The
flux between 0.5 and 2.4 keV is $(5.06,5.69,10.33,6.54)\times
10^{-11}\rm\ ergs\,cm^{-2}s^{-1}$, respectively. All the models
include a Galactic absorption fixed at $N_H = 1.36\times10^{20}\rm\
cm^{-2}$ (Lockman \& Savage~1995; errors are quoted at 90\%
confidence). 

\begin{figure*}
\begin{center}
\includegraphics[angle=90,width=0.9\textwidth,height=0.6\textheight]{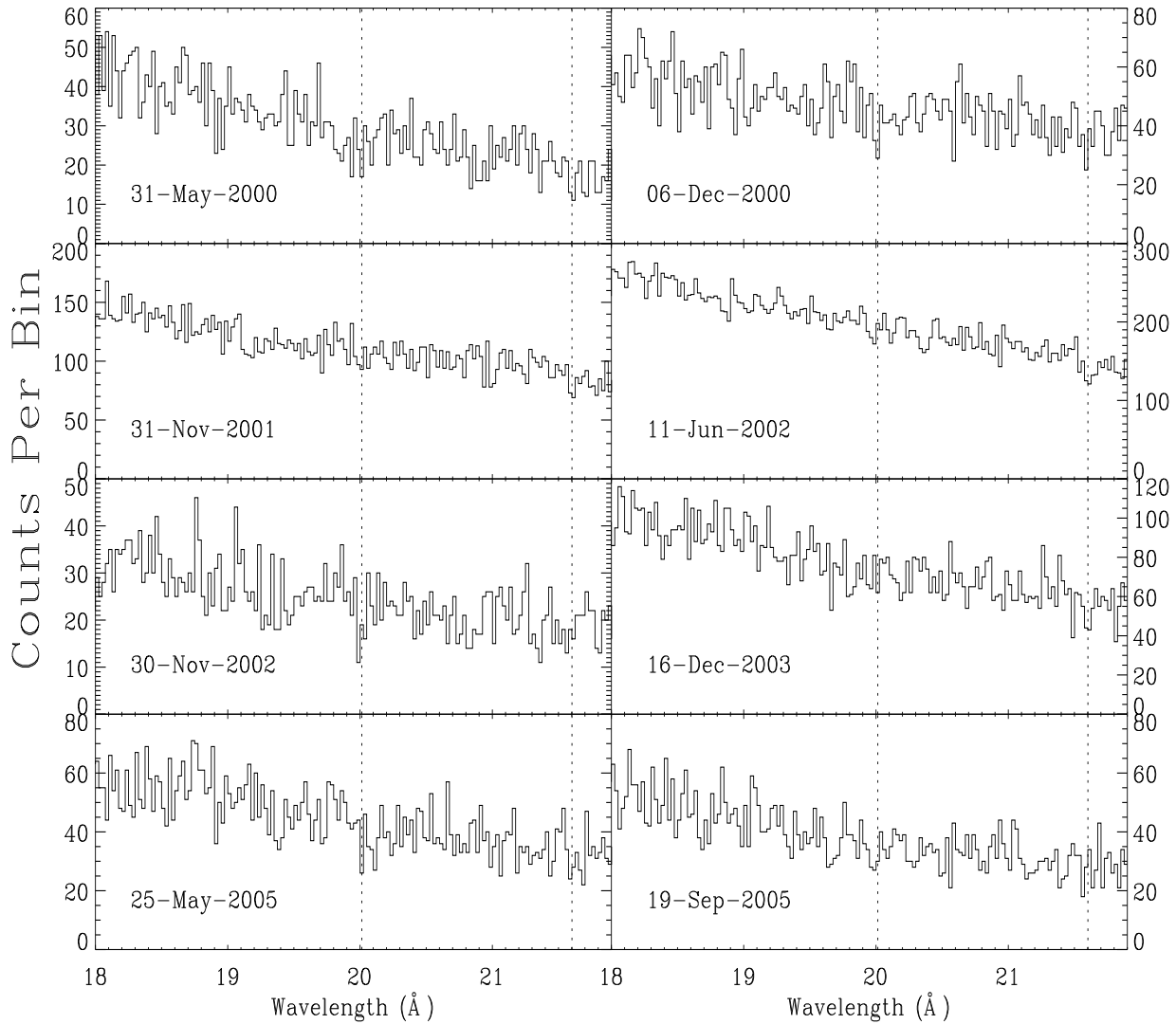}
\caption{Raw counts for the five observations between 18 and 22
  \AA. Bin size is 0.025 \AA.\ Two vertical dashed lines indicated positions of the redshifted \ion{O}{8} and
  rest \ion{O}{7} lines from FANG02.}
\label{fig:total1}
\end{center}
\end{figure*}

In Figure~\ref{fig:total1} we show the raw counts of the five
observations. The bin size is 0.025 \AA.\ The line spread function (LSF) of LETGS has a typical full width of half maximum (FWHM) of $\sim 0.05 \AA\ $. In this way one FWHM will contain two bins. Two vertical dashed lines indicate the wavelength of the redshifted \ion{O}{8} line (at $\sim 20$
\AA) reported by FANG02, and rest \ion{O}{7} lines (at $\sim 21.6$ \AA). There are no
strong absorption lines in these observations, but we do find
small dips in each of these observations at the corresponding positions.

\begin{figure*}
\begin{center}
\vskip-1.5cm
\resizebox{6in}{!}{\includegraphics[angle=90,width=1.\textwidth,height=0.7\textheight]{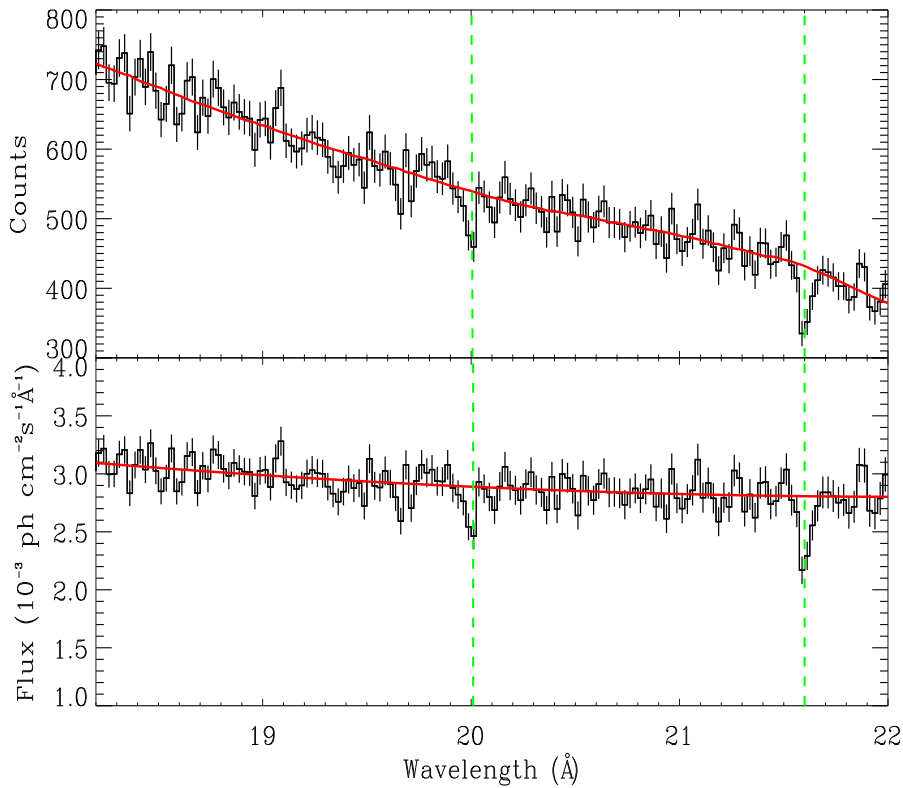}}
\end{center}
\vskip-1.5cm
\caption{The top panel shows the raw counts of the stacked data, with a binsize of 0.025 \AA.\ The bottom panel shows the flux spectrum. The red line show the best
  fitted model. Two vertical green dashed lines indicate positions of
  the redshifted \ion{O}{8} and rest \ion{O}{7} lines from FANG02.} 
\label{fig:total2}
\end{figure*}

To enhance the signal we stack all the eight data sets. To ensure that
all the eight observations are properly aligned with each other, we
also compare the wavelengths of two known features, atomic \ion{O}{1}
from our Galaxy and solid-state \ion{O}{1} from the instruments, and
calibrate the five data sets against these two lines
(FANG02). Only two observations (1703 and 2335) showed a systematic
shift of $\sim 0.05$ \AA\ shift, and we have to adjust the shift
manually. Figure~\ref{fig:total2} shows the stacked data between 18
and 22 \AA.\,We have achieved from $\sim$ 450 to 600 counts per bin
here, compared to $\sim 100$ counts per bin in FANG02. We again mark
the positions of the two absorption lines.

Measurement of narrow absorption features depends crucially on
determining the correct continuum level. However, a simple power law
cannot provide the best fit to the continuum of the stacked data for:
1) there are spectral variations among different observations, and 2)
there are residual uncertainties in instrumental efficiencies. Instead we fit the local continuum using polynomials. More
specifically, we first fit the stacked data with a single power law
plus neutral hydrogen absorption as an initial guess. The residuals
between 6 and 42 \AA\, are then fitted with a six-order
polynomial. Such a polynomial will get rid of any features with a
characteristic scale of $\gtrsim 6$ \AA, but smaller features will be
preserved. Furthermore, to account for small scale fluctuations, we
divide the regions between 6 and 42 \AA\, into 9 bands, with a
bandwidth of 4 \AA\, each. For each region we then fit the residuals
from the previous power law fitting with a polynomial. The order of
this polynomial is given by the one that has the smallest
$\chi^2$. The red lines in Figure~\ref{fig:total2} show the final
result of the fitted continuum in the 18 -- 22 \AA\ range. The top panel shows the spectrum in units of counts per bin, with a binsize of 25 m\AA, and the bottom panel shows the flux,  in units of $\rm photons\ cm^{-2}s^{-1}\AA^{-1}$.

\begin{figure}[t]
\begin{center}
\resizebox{3.5in}{!}{\includegraphics[angle=90,width=0.4\textwidth]{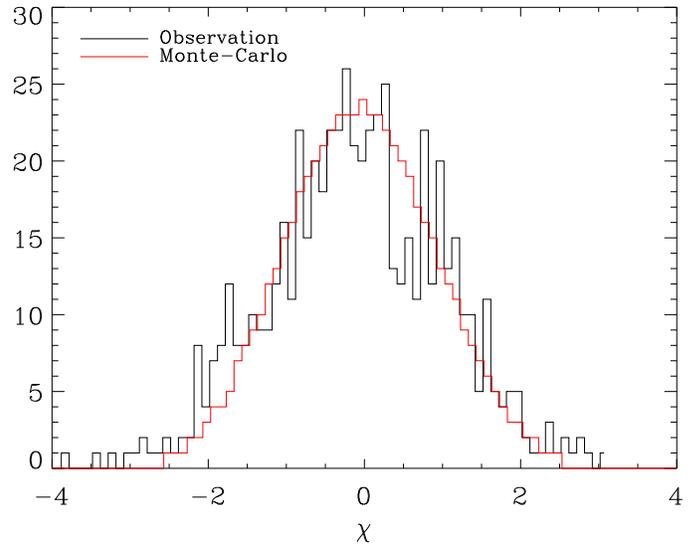}}
\end{center}
\caption{$\chi$-distribution of the data (dark line) and of 10,000
  Monte-Carlo simulations (red line).} 
\label{fig:resid}
\end{figure}

How well does our model depict the observed continuum? We first
calculate the $\chi$-distribution of the data. $\chi$ is defined as:
\begin{equation}
\chi = \frac{data\ count - model\ count}{(model\ count)^{1/2}}.
\end{equation}
We then run 10,000 Monte-Carlo simulations, each simulation is a
random realization of the model based on Poisson statistics, and we
also calculate the $\chi$-distribution of these
simulations. Figure~\ref{fig:resid} shows the comparison between the
data (dark line) and the simulation (red line). We find in general the
$\chi$-distribution of the data follows that of the simulation
reasonably. The data does show a long, negative tail,
this is actually caused by the absorption features that we will
discuss below. Excluding the three prominent absorption features
that we will discuss in the next section, we perform a
Kolmogorov-Smirnov test on the two distributions. The null hypothesis
that the data sets are drawn from the same distribution has a
significant level of $\sim 0.48$, suggesting that our model fits the
continuum quite well, and that our fitting procedure does take care of
small scale fluctuations. 

\begin{table}
%\footnotesize
\large
\caption{~~~~~~~~Line Fitting Parameters~~~~~~~~}
\begin{center}
\begin{tabular}{lllll}
\hline
\hline
                       & & \ion{O}{8} Ly$\alpha$\tablenotemark{a} & &  \ion{O}{7} He$\alpha$\tablenotemark{a} \\
\hline
$\rm Wavelength$       & & $20.00\pm0.01$            & & $21.60\pm0.01$ \\
$cz({\rm km\ s^{-1}})$ & & $16307\pm158$             & & $-26\pm138$ \\
Line Width\tablenotemark{b}       & & $< 0.027$                   & & $<0.029$ \\
$\rm Line\ Flux$\tablenotemark{c}   & & $2.1_{-0.8}^{+0.6}$         & & $3.9_{-1.1}^{+0.8}$ \\
EW (m${\rm \AA}$)      & & $7.42_{-1.94}^{+2.76}$        & & $13.75_{-2.80}^{+4.09}$ \\
SNR\tablenotemark{d}              & & 5.0                         & & 7.1 \\
\hline
\end{tabular}
\end{center}
\noindent a. Rest-frame wavelengths for \ion{O}{8} Ly$\alpha$ and \ion{O}{7}
He$\alpha$ are 18.9689 \AA\, and 21.6019 \AA\, respectively (Verner et
al.~1996). Errors are quoted at 90\% confidence level hereafter.\\
\noindent b. 90\% upper limit of the intrinsic line width $\sigma$, in units of
\AA.\ The FWHM of the fitted Gaussian line is $2.35\sigma$.\\ 
\noindent c. Absorbed line flux in units of $\rm 10^{-5}~photons~cm^{-2}s^{-1}$.\\
\noindent d. In units of equivalent sigma of a Gaussian distribution with the
same confidence level as the Poisson significance gives, based on the $\chi$-distribution.\\
\end{table}

\section{Results}

\begin{figure*}[t]
\begin{center}
\resizebox{6in}{!}{\includegraphics[angle=90]{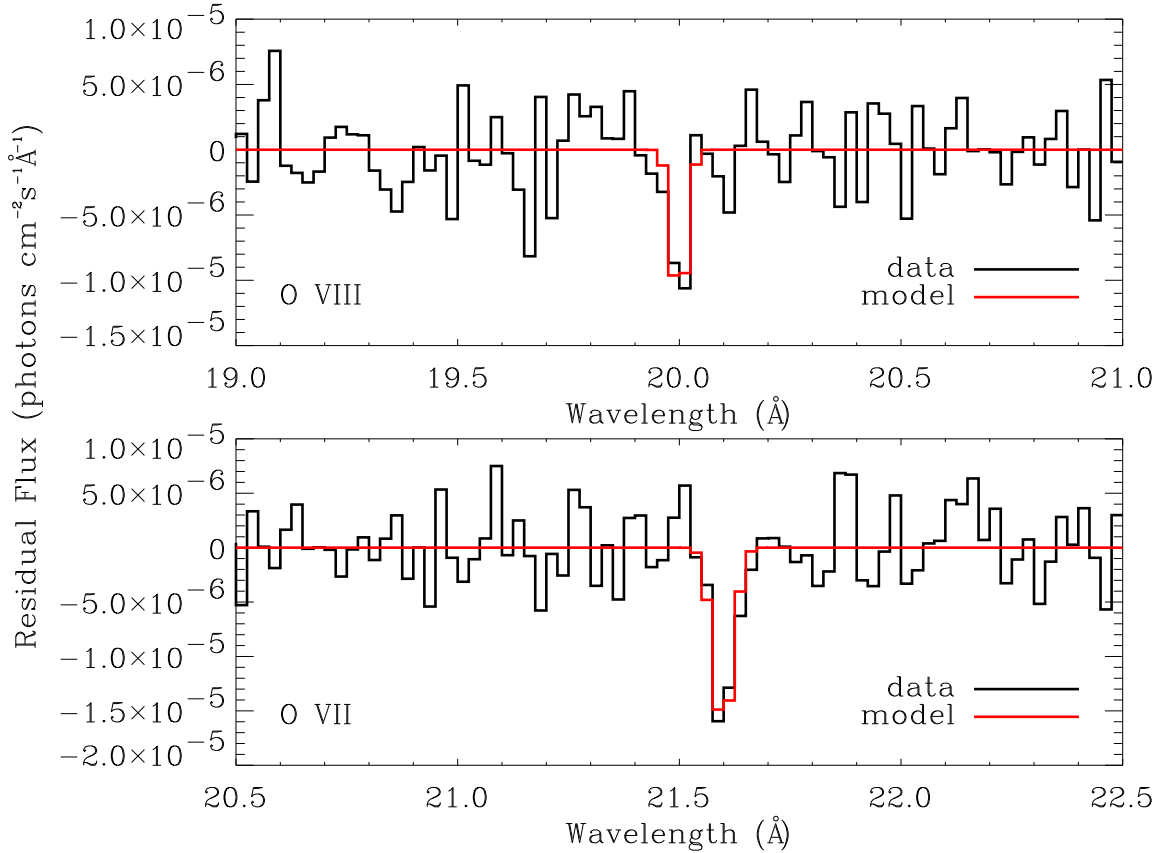}}
\end{center}
\caption{Residual flux after subtracting the best-fitted continuum
  from the stacked data. The red lines show the ISIS-fitted Gaussian
  line profiles. The top panel is for the redshifted \ion{O}{8} line
  and the bottom panel is for the zero-redshift \ion{O}{7} line.}
\label{fig:lines}
\end{figure*}

A blind search of the entire region from 6 to 42 \AA\ for features at $>3\sigma$ level give only three prominent absorption features, at $\sim 20.0$, 21.6, and 23.5 \AA.\, The last one is an interstellar \ion{O}{1} line and will not be discussed here. These two features at 20.0 and 21.6 \AA\ are subsequently analyzed with customized routines from the software package ISIS (Interactive Spectral
Interpretation System, see Houck \& Denicola~2000) \footnote{see
  http://space.mit.edu/ASC/ISIS/}. Specifically, we subtract the data
from the best-fit model, and then fit the residual with a Gaussian
line profile. The EW is an integration of $\left(1-I/I_0\right)$,
where $I$ is the observed spectrum, and $I_0$ is the continuum.  We
list the fitting parameters in Table~2. The Gaussian line fitting
gives a line width ($\sigma = 0.017$ and $0.021\rm\ \AA$ for the 20.0 and 21.6 \AA\
lines, respectively) that is even narrow than that of the instrumental
LSF, we only present the 90\% upper limits on the
intrinsic line widths. In calculating SNR (signal-to-noise ratio),
signal is the
total photon counts missed in the line, and noise is square-root of continuum
photon counts within the line profile. The continuum photon counts are the total of counts
under continuum, with a width of three times the measured line width, slightly
larger than the FWHM of the fitted Gaussian line. In
Figure~\ref{fig:lines} we show the residual spectra of both features
(black lines) and the fitted Gaussian lines (red lines). 

\subsection{Contamination from the Telescope Features?} 

Before we proceed to further discussion, we need to understand the effects of any fixed-position detector features (such as those chip gaps and node boundaries) and instrumental features (such as those absorption edges in the telescope materials). While over the entire wavelength range the telescope has been well calibrated, it is still likely that some of the variations of the effective area and/or detector quantum efficiency across these features can affect the estimated significances of our detected lines, if it happens that the detected lines are close to these telescope features.

We do not find any instrumental features near our two detected lines
(POG\footnote{{\sl Chandra} Proposers' Observatory Guide, see
  http://asc.harvard.edu/proposer/POG}, Table~9.4). The positions of
detector features depend on the aimpoint of each observation. For
commonly used offset pointing ($\Delta y = 1.5\arcmin$ and $\rm SIM-Z
= -8 mm$), the nominal aimpoint in S3 chip is moved from
$\rm(chipx,chipy)=(231,502)$ to $\rm(chipx,chipy)=(49,168)$, which
also avoids the zeroth-order being at one of the node boundaries. A
careful calculation indicates no telescope features around the 21.6
\AA\ line, but such configuration does put the positive first order
20.0 \AA\ line close to the boundary between node 2 and node 3 in S3
chip. However, we believe this is unlikely to have large impact on the
significance of the 20.0 \AA\ line because of the following
considerations. First, as we stated before, Obs.\#3669, the one with
the most photon counts (accounts for almost half of the total counts
at 20.0 \AA), had a $y$-offset of $3.3\arcmin$, which put the aimpoint
in chip S2. This results in the positive first order 20.0 \AA\ line
being far away from any telescope features. Secondly, this line is
clearly visible in both positive and negative first order spectra. It
is unlikely that this line would be detected in the negative order if
it is caused by detector feature in the positive order. Thirdly, and
most importantly, we believe that our continuum-extraction technique
can take care of the small variation caused by node boundaries. During
most observations, the spacecraft is dithering to (1) smear out the
reduced exposure from chip gaps, and (2) smooth out pixel-to-pixel variation in
the response (see POG). The standard dithering pattern has a
peak-to-peak width of $16\arcsec$, this roughly corresponds to a
response variation on the scale of $\rm \sim 1 \AA\ $ for
LETGS. However, based on the fitting procedures we described in
section \S3, the residual between 18 and 22 \AA\ , after broadband
continuum-fitting with a power law and 6-order polynomial, was fitted
with another 6-order polynomial. This would get rid of any feature
with a typical scale of $\rm \gtrsim 0.7\ \AA\ $, which includes the
dithered node boundary. The effect of such procedure can be viewed from
Figure~\ref{fig:total2}: in the top panel, the raw counts spectrum shows
some variation between 19 and 21 \AA\ , however, in the bottom panel, such
variation has been smoothed out in the flux spectrum.    

\subsection{Comparison with results from FANG02}

Most importantly, our work does detect an absorption line at
5$\sigma$ significance at $\sim 20$ \AA,\ confirming the initial
detection by FANG02. We find that the results presented in this paper
are consistent with those of FANG02, but the estimated EW in this work
($7.42_{-1.94}^{+2.76}$) is about one-half of that estimated
earlier, based on many fewer counts (150 vs.  550 counts per bin) and with larger uncertainty. The two results
are consistent at 90\% confidence level. The properties of the
stronger local ($z
\approx 0$) \ion{O}{7} absorption line are fully consistent with those reported in FANG02.  

\subsection{Comparison with results from {\sl Chandra} LETG+HRC-S}

For a consistency check, we also analyzed the observations 
up to a date of 2004 November 22 taken with the LETG operated with High Resolution Camera for 
spectroscopy (HRC-S; Table~3). We reprocessed these observations
using the {\sl Chandra} Interactive Analysis of Observations (CIAO) software 
(ver. 3.3.0.1) with the calibration database CALDB (ver. 3.2.1). Because the
HRC itself does not have intrinsic energy resolution, the photons located 
to the same position of the grating arms but from 
different grating orders cannot be sorted out as in those observations with
ACIS. We therefore used a forward approach to account for the spectral order 
overlapping. For each observation, we first calculated the response matrix 
files (RMFs) and the auxiliary response functions (ARFs) for grating orders 
from the first to sixth (assuming the contributions to the spectrum from the 
higher $>6$ orders are negligible). We then added these six RMF and ARF pairs
to form an order-combined response (RSP) file using IDL routines from 
PINTofALE package\footnote{http://hea-www.harvard.edu/PINTofALE}.
To enhance the counting statistics, spectra from all the HRC-S observations,
after a consistency check, were co-added, and the RSP files are 
weight-averaged based on the corresponding exposure time and the continuum 
intensity obtained from a global fit to each individual spectrum.

\begin{table}
%\footnotesize
\large
\caption{HRC Observations}
\begin{center}
\begin{tabular}{rcc}
\hline
\hline
ObsID & Date & Exposure (ks) \\
\hline
331   & 1999 Dec. 25 & 63.2 \\
1013  & 2001 Apr. 06 & 26.8 \\
1704  & 2000 May  31 & 26.0 \\
3166  & 2001 Nov. 30 & 30.0 \\
3709  & 2002 Nov. 30 & 13.8 \\
4406  & 2002 Nov. 30 & 13.9 \\
5172  & 2004 Nov. 22 & 27.2 \\
\hline
\end{tabular}
\end{center}
\end{table}

\begin{figure}[t]
\begin{center}
\resizebox{3.in}{!}{\includegraphics[angle=270]{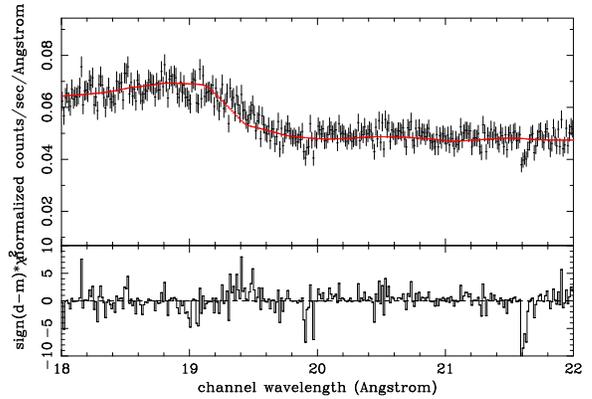}}
\end{center}
\caption{LETG+HRC spectrum. The red line in the top panel is the model. The bottom panel shows $\chi^2$ with a sign.}
\label{fig:hrc}
\end{figure}

In Fig.~\ref{fig:hrc} we show the total HRC spectrum (from the first
to sixth order) between 18 and 22 \AA.\ The 21.6 \AA\ line is clearly
visible, but not the 20.0 \AA\ line. There are some hints of
absorption features between 19 and 20 \AA,\ but nothing near the
LETG+ACIS detected 20.0 \AA\ line. We cannot extract the first order
spectrum because of the reason we described above, but using a first
order response matrix, we estimate the continuum flux of the first
order is about 90\% of the total flux, which gives $\sim$ 450 counts
per 0.025 \AA\ at $\sim$ 20.0 \AA\ region. The observations with the
LETG+ACIS have a slightly stronger continuum ($\sim$ 550 counts per
0.025 \AA\ at the same wavelength). We estimate the upper limit of an
absorption line EW, by adding such line in the spectrum and
calculating $\Delta\chi^2$ by varying line EW. This gives an upper
limit of $\sim 9$ m\AA,\ which is consistent with the result from the
LETG+ACIS observations. 

To further investigate the difference between the results from the
LETG+ACIS and LETG+HRC, we compare their LSFs. In
Figure~\ref{fig:lsf}, we show the LSFs of the LETG+ACIS (red line)
and the LETG+HRC (green line) at 20 \AA.\ The blue line is the LSF
of the {\sl XMM} RGS1, and we will discuss it later. Clearly, the HRC curve has a rather broad wing. This
suggests that when compared with that of the ACIS, more photons will
be distributed in the wing, and the significance at the line center
will be lower. Such broader profile is likely caused by the
non-linearities of the LETG+HRC-S dispersion relation\footnote{For
  detailed discussion on the non-linearities of the LETG+HRC-S
  dispersion relation, see
  http://cxc.harvard.edu/cal/Letg/Corrlam/.}. To test the impact of the LSF, we run 10,00 Monte-Carlo
simulations for both the LETG+ACIS and the LETG+HRC, based on the
Poisson statistics. Each simulation is a random realization of the
corresponded continuum model plus an absorption line at 20 \AA.\ We
adopt line parameters from the \ion{O}{8} line in Table~2 with an
intrinsic line width of 0.016 \AA,\ and then
fold it through the corresponded LSF. We find that out of 10,000
trials, the line is detected at $3\sigma$ level or higher in more than half of
the ACIS simulated spectra ($\sim$ 57\%); however, for the HRC
spectra, the detection probability decreases to about one-fourth
($\sim$ 25\%).  

\subsection{Comparison with results from {\sl XMM}-Newton}

PKS~2155-304 was also extensively observed with {\sl
  XMM}-Newton. A detailed spectroscopy study of the RGS data was
  presented by Cagnoni et al.~(2004). They confirm the detection of
  zero-redshift \ion{O}{7} absorption line at a significance level of
  4.5$\sigma$. Their best fit equivalent width
  ($EW=19.50_{-8.17}^{+7.89}$ m\AA) is higher than what we measure but
  consistent at 90\% confidence level. They
  also claim that this line profile is possibly double
  peaked. Although the {\sl Chandra} LETG-ACIS has higher energy resolving power, we cannot confirm this structure in our data.

Cagnoni et al.~(2004) stated that they do not detect the redshifted \ion{O}{8}
absorption line at $\sim 20$ \AA,\ reported by FANG02, and set a
3$\sigma$ upper limit of 14 m\AA.\ However, this value is large enough to be consistent with both the improved measurement and the original best fit value of FANG02 for the detected line.

To further compare the {\sl Chandra} and {\sl
  XMM}-Newton data, we reanalyze the {\sl XMM}-Newton data presented
  in Cagnoni et al.~(2004) with the same techniques adopted in this
  paper. In doing this we hope we can minimize the differences that
  can be caused by various data analysis techniques. For instance,
  whereas we apply a adapted-polynomial plus power law fit technique
  to subtract continuum, they use local absorbed power law to fit
  individual regions with bandwidths of 2 -- 3 \AA.\,

We use the three data sets that were analyzed in Cagnoni et al.~(2004) (ObsID
\#0080940101, 0080940301, and  0080940401). We do not select the
fourth observation (ObsID \#0080940501) because it contains a
background flare. These three observations have a total exposure time
of $\sim 110\,ksec$ (see Table~1 of Cagnoni et al.(2004) for detailed
observation log). We analyze the data with the standard Science
Analysis System (SAS) version 5.4.1 \footnote{See
http://xmm.vilspa.esa.es/}. We use only RGS-1 data since there is no
data between 20 and 24 \AA\, from RGS-2 due to the failure of a CCD
chip. The LSF of the RGS-1 has a FWHM of $\sim
0.06$ \AA\ with rather extended wings \footnote{See {\sl XMM}-Newton Users' Handbook at
http://xmm.vilspa.esa.es/.}. We will not present our data extraction
procedures here, but refer reader to Cagnoni et al.~(2004) for details. After
obtaining the first order spectrum, we follow procedures that are
described in section \S3 to subtract the
continuum. Figure~\ref{fig:xmm_line} shows the RGS-1 spectrum between
19.7 and 20.7 \AA.\ We obtain $\sim 450$ counts per 0.025 \AA\
bin. This is 20\% lower than what we observed in {\sl Chandra} data ($\sim
550$ counts per 0.025 \AA\ ).

The dark solid line in figure~\ref{fig:xmm_line} shows the {\sl XMM}
data between 19.5 and 20.5 \AA.\, No absorption line is seen around
$\sim$ 20 \AA. With our continuum subtraction techniques, we obtain a
3$\sigma$ upper limit of $\sim 10$ m\AA,\ consistent with what is
obtained in Cagnoni et al.~(2004) and the $\sim 7$ m\AA\ equivalent width line
we report in this paper. To demonstrate this, in
Figure~\ref{fig:xmm_line} we plot a 3$\sigma$, 10 m\AA\ line at $\sim$
20 \AA\, in green. An illustration (but without imposing statistical
fluctuations) of how the line detected in {\sl Chandra} would appear
in the {\sl XMM} data is plotted in red. We adopt our {\sl Chandra}
line parameters, and then convolve the line with the RGS1
LSF\footnote{A template of the LSF was kindly provided by A.~Rasmussen.}. 

\begin{figure}[t]
\begin{center}
\resizebox{3.in}{!}{\includegraphics[angle=0]{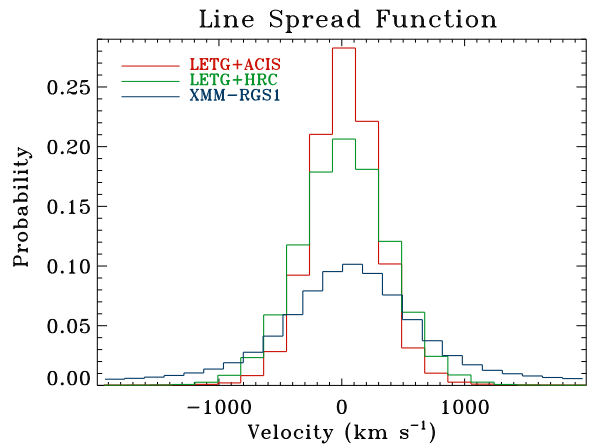}}
\end{center}
\caption{The line spread functions of the {\sl Chandra} LETG+ACIS
  (red), the {\sl Chandra} LETG+HRC (green), and the {\sl XMM}
  RGS1(blue) at 20 \AA.}
\label{fig:lsf}
\end{figure}

In Figure~\ref{fig:xmm_ewchi} we show the $\Delta \chi^2$ of fitting
the {\sl XMM} data with a continuum plus a Gaussian line model (again,
convolved by the LSF), where $\Delta \chi^2 = 0$ is defined as no
Gaussian line. We fix the line center at $\lambda = 20$ \AA,\, and
gradually vary the line equivalent width. The horizontal dashed line indicates the $\Delta \chi^2 = 9$,
or 3$\sigma$ confidence. Three vertical dark lines show the line width
and 90\% lower and upper limits of our detection. Again, we find that
one would not have expected {\sl XMM}-Newton to detect the line seen
with {\sl Chandra}.

Two main reasons that explain the non-detection of the X-ray absorption
line in the {\sl Chandra} LETG+HRC data may also apply there. First, the continuum is slightly higher in {\sl Chandra} data. At 20 \AA,\ the {\sl Chandra} observations show $\sim 550$
counts per bin, compared with $\sim 450$ counts in the {\sl XMM}
data. Secondly, and most importantly, the LSF of the RGS1 not only is
larger than that of the LETGS, but also has rather extended wings. As
pointed out by Williams et al.~(2006) in a study of the {\sl XMM} data
of Mrk~421, the central 0.1 \AA\ region of a line contains about 96\%
of the total line flux for {\sl Chandra} LETG, for RGS this number
decreases to $\sim 68\%$. In Figure~\ref{fig:lsf} the dotted line is
the LSF of the RGS1. The wings of the RGS1 are even broader than that
of the LETG+HRC. We ran a similar Monte-Carlo simulation on the RGS1
data.  The result supports our finding: out of 10,000 trials, a
$3\sigma$ line is detected in only $\sim$ 10\% RGS1 spectra, compared
with nearly $\sim$ 57\% detection in the LETG+ACIS spectra, and $\sim$
25\% detection in the LETG+HRC spectra. 

\section{Discussion}

Following FANG02 and assuming that the line is not saturated, the
column density of the redshifted \ion{O}{8} absorption line at $20$
\AA\, is N(\ion{O}{8})$ = 5.0^{+1.9}_{-1.3}\times
10^{15}\rm\,cm^{-2}$. The line width sets an upper limit to the path
length of $\sim 5.8\,h^{-1}_{70}$ Mpc based on the Hubble flow
\footnote{We use $H_0 = 70\,h_{70}\rm\,km\,s^{-1}Mpc^{-1}$, and a
standard $\Lambda$CDM model with $\Omega_m=0.3$ and $\Omega_{\Lambda}
= 0.7$ throughout the paper}. This gives $n_{b} >
5\times10^{-6}\,\rm cm^{-3}$ $Z_{0.1}^{-1}f_{0.5}^{-1}l_{5.8}^{-1}$
where $Z_{0.1}$ is the metallicity in units of 0.1 solar abundance
\footnote{We adopt solar abundance from Anders \& Grevesse~(1989)}, $f_{0.5}$ is
the ionization fraction in units of 0.5 and $l_{5.8}$ is the path
length in units of $5.8\,h_{70}^{-1}$ Mpc. A lower limit to the path
length of $\sim$ 1 Mpc can be obtained by assuming the absorber has
the size of the small galaxy group detected in 21 cm images (Shull et
al.~1998), which gives $n_{b} \approx 3.1\times
10^{-5}\rm\,cm^{-3}$ $Z_{0.1}^{-1}f_{0.5}^{-1}$.  This implies a range of
baryon overdensities of $\delta_b \approx 30 - 150$, which, as found
in FANG02, is consistent with the predicted density of the WHIM gas
from cosmological hydrodynamic simulations (see, e.g., Cen \&
Ostriker~1999; Dav\'{e} et al.~2001).

To constrain the temperature we need to study the ionization structure
of the \ion{O}{8} absorber. Assuming collisional ionization equilibrium, the
ionization fraction is a function of temperature only, and
$f$(\ion{O}{8}), the ionization fraction of \ion{O}{8}, has a peak
value of $\sim$ 0.5 between 2 and 5 $\times 10^6$ K. A constraint on
the temperature of the \ion{O}{8} absorber can be obtained by studying the
column density ratio between \ion{O}{8} and \ion{O}{7}. Non-detection
of the \ion{O}{7} at the corresponding redshifted position
($\lambda=22.77$ \AA) with a 3$\sigma$ upper limit on the
equivalent width of $\sim$ 9.6 m\AA,\,implies N(\ion{O}{7}) $ \lesssim
3\times 10^{15}\rm\,cm^{-2}$, or
$\log[$N(\ion{O}{8})/N(\ion{O}{7})$] \gtrsim 0.22$. This constrains
the temperature to be $T \gtrsim 2.3\times 10^6$ K. However, we need
to be careful about the assumption of collisional ionization
equilibrium (Cen \& Fang~2006). 

\begin{figure}[t]
\begin{center}
\resizebox{3.5in}{!}{\includegraphics[angle=90]{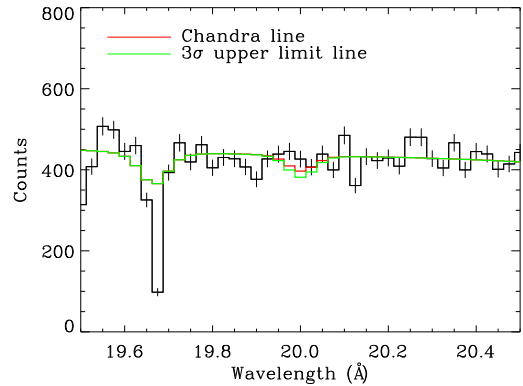}}
\end{center}
\caption{{\sl XMM} data at $\sim 20$ \AA.\, The dark line shows the
  data, the green line shows the continuum with a 3$\sigma$ absorption
  line, and the red line shows the continuum with the line detected
  in {\sl Chandra} data. The line at $\sim$ 19.75 \AA\ is an instrumental feature.}
\label{fig:xmm_line}
\end{figure}

We can actually strengthen the constraint on temperature by searching
for potentially co-existing absorption lines at other
wavelengths. Recently, Shull et al.~(2003) reported the detection of a
series of \ion{H}{1} and \ion{O}{6} absorbers along the sight line
towards PKS~2155-304 between 16185 and 17116 $\rm\,km\,s^{-1}$. Based
on FANG02 they resorted to a multi-phase model to explain the
kinematic offsets among \ion{H}{1}, \ion{O}{6} and \ion{O}{8}
absorbers. The new measurement presented here indicates that these
absorbers may coexist owing to the small separation in velocity
space. For instance, the X-ray absorber may coexist with the component
A in their observation, which is located between 16185 and 16252
$\rm\,km\,s^{-1}$ and shows \ion{O}{6} absorption line. If this is
indeed a single absorber, we can set a tight range of the
\ion{O}{6}-to-\ion{O}{8} ratio:  $-2.68 \lesssim
\log[$N(\ion{O}{6})/N(\ion{O}{8})$] \lesssim -1.90 $. This in turn
sets temperature: $1.6 \lesssim T \lesssim 2.3 \times 10^6$ K. Using
the previous constraint from the \ion{O}{8}-to-\ion{O}{7} ratio, we
can put a tight constraint on the temperature of the component A: $T \approx 2.3 \times 10^6$ K.

\begin{figure}[t]
\begin{center}
\resizebox{3.5in}{!}{\includegraphics[angle=90]{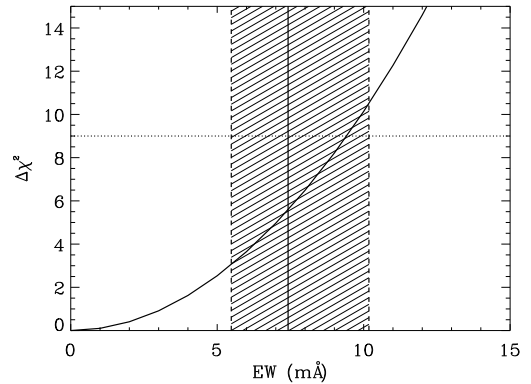}}
\end{center}
\caption{$\Delta \chi^2$ from fitting {\sl XMM} data with a continuum
  plus a Gaussian line. The horizontal line shows the 3$\sigma$ upper
  limit. The shadowed red area indicates the equivalent width and
  90\% confidence range of the line that is detected in {\sl Chandra}
  data.}
\label{fig:xmm_ewchi}
\end{figure}

Our discussion on the density and temperature of the 20.0 \AA\
absorber depends on several crucial assumptions that need to be
closely examined. First, we assume the line is unsaturated. However,
the 90\% upper limit on the intrinsic line width can only give a
Doppler-$b$ parameter of $\lesssim 566 \rm\ km\ s^{-1}$, much
larger than the typical thermal broadening width $\sim 50\rm\ km\
s^{-1}$of a few million-degree gas. The non-detection of the high-order Lyman series, particularly the
Ly$\beta$ line at the rest-frame of $\sim 16$ \AA,\ can rule out the
saturate scenario at $1\sigma$ level only, and we notice in some cases
the local X-ray absorption line could be saturated (Williams et
al.~2006; Yao \& Wang~2006, 2007). If the line is saturated, the derived
column density can be treated as a lower limit only. If the line is
indeed unsaturated, it will be broad. As suggested by the clump-infall
model in Shull et al.~(2003), a shock-wave of $\sim 400 \rm\ km\ s^{-1}$
from structure evolution can provide such a non-thermal broadening
mechanism: such a shock-wave would produce a post-shock temperature of
$\sim 2\times 10^6$ K.

Secondly, we assume the gas is in collisional
ionization equilibrium. Recent study, using numerical simulation,
shows that non-equilibrium evolution of various ion species can
produce significantly different effects (Cen \& Fang~2006; Yoshikawa
\& Sasaki~2006), since the timescales for ionization and recombination are
not widely separated from the Hubble timescale. Also, the relaxation
process between electrons and ions may produce a two-temperature
structure in the IGM and have impact on the metal ionization fraction,
since the relaxation time is comparable to the Hubble timescale
(Yoshida et al.~2005). However, we expect such process would be more
important for hotter gas at $T>10^7$ K (Yoshida et al.~2005). At
very low density, especially at $n_b \lesssim 10^{-5}\rm\,cm^{-3}$,
photoionization can also become important (see, e.g., Hellsten et al.~1998;
Nicastro et al.~2002; Chen et al.~2003; Williams et al.~2005). For instance, at $n_b =
10^{-6}\rm\,cm^{-3}$, the \ion{O}{8} ionization curve peaks at $\sim
3\times10^5$ K under the cosmic X-ray background radiation (Chen et al.~2003). 

Finally, we discuss the coexistence of \ion{O}{6} and
\ion{O}{8} absorbers primarily because of the proximity in radial
velocities. However, as suggested by Shull et al.~(2003), the \ion{O}{6}
line width is narrow ($40 \pm 10\rm\ km\ s^{-1} $ FWHM), and the
scenario that \ion{O}{6} and \ion{O}{8} can coexist is marginally
plausible within measurement errors ($\log T \approx 6.25 \pm 0.1$).    

The baryonic content $\Omega_b$(\ion{O}{8}) that is probed by the
\ion{O}{8} absorption can be estimated following Lanzetta et
al.~(1995) and Tripp et al.~(2000). Given the path length of the sight line toward
PKS~2155-304 of $\Delta z \approx 0.116$ and assuming an upper limit
of metallicity of $0.5Z_{\odot}$, we estimate $\Omega_b$(\ion{O}{8})
$\gtrsim 0.004\,h^{-1}_{70}$, or about 10\% of the total baryon
fraction. This number is consistent with the prediction of the WHIM
gas from numerical simulations. Based on this single detection, the
observed distribution $dn/dz$, defined as number of absorbers per unit
redshift, is higher than what simulation predicts (see, e.g., Cen \&
Fang~2006, Fig.~4).  We certainly need more detections to improve
statistics. 

Note: After we submitted this paper, Williams et al.~(2006b) published
a paper on the {\sl Chandra} archival data of PKS~2155-304. While they
mainly focused on the local $z=0$ absorption lines, they also
discussed the intervening absorption line we reported in FANG02. They
confirmed the detection of this line using the {\sl Chandra}
LETG-ACIS. The measured equivalent width is ($7.5\pm2.1$ m\AA)\ is
consistent with what we found, although the significance is lower
($3.5\sigma$). They did
not detect this line using the {\sl Chandra} LETG-HRC, but obtained an
upper limit of $\sim 12.5$ m\AA.\ This is in agreement with what we
found with the HRC data, and is consistent with the detection in the
ACIS data.

\acknowledgments
We thank Herman Marshall for providing helpful
IDL tools for data analysis. We also thank the referee Fabrizio
Nicastro for useful
suggestions. We thank David Buote and Andrew Rasmussen for useful discussions. TF was supported by the NASA through {\sl
Chandra} Postdoctoral Fellowship Award Number PF3-40030 issued by the
{\sl Chandra} X-ray Observatory Center, which is operated by the
Smithsonian Astrophysical Observatory for and on behalf of the NASA
under contract NAS 8-39073. CRC and YY are supported by NASA through the Smithsonian Astrophysical
Observatory (SAO) contract SV3-73016 to MIT for support
of the Chandra X-Ray Center, which is operated by the SAO
for and on behalf of NASA under contract NAS 08-03060. YY is also
supported by AR7-8014.

\end{document}